# Reemergence of high-$T_c$ superconductivity in the $(Li_{1-x}Fe_x)OHFe_{1-y}Se$ under high pressure

J. P. Sun,[1,2,=] P. Shahi,[1,=] H. X. Zhou,[1,2,=] Y. L. Huang,[1,2] K. Y. Chen,[1,2] B. S. Wang,[1,2] S. L. Ni,[1,2] N. N. Li,[3] K. Zhang,[3] W. G. Yang,[3,4] Y. Uwatoko,[5] K. Jin,[1,2] F. Zhou,[1,2] D.J. Singh,[6] X. L. Dong,[1,2*] Z. X. Zhao,[1,2] and J.-G. Cheng[1,2*]

[1]Beijing National Laboratory for Condensed Matter Physics and Institute of Physics, Chinese Academy of Sciences, Beijing 100190, China

[2]University of Chinese Academy of Sciences, Beijing 100049, China

[3]Center for High Pressure Science and Technology Advanced Research (HPSTAR), Shanghai 201203, China

[4]High Pressure Synergetic Consortium (HPSynC), Geophysical Laboratory, Carnegie Institution of Washington, 9700S Cass Avenue, Argonne, Illinois 60439, USA

[5]The Institute for Solid State Physics, University of Tokyo, Kashiwa, Chiba 277-8581, Japan

[6]Department of Physics and Astronomy, University of Missouri, Columbia, Missouri 65211-7010, USA

E-mails: jgcheng@iphy.ac.cn or dong@iphy.ac.cn

## Abstract

The pressure-induced reemergence of the second high-$T_c$ superconducting phase (SC-II) in the alkali−metal intercalated $A_xFe_{2-y}Se_2$ (A = K, Rb, Cs, Tl) remains an enigma and proper characterizations on the superconducting- and normal-state properties of the SC-II phase were hampered by the intrinsic inhomogeneity and phase separation. To elucidate this intriguing problem, we performed a detailed high-pressure magnetotransport study on the recently discovered $(Li_{1-x}Fe_x)OHFe_{1-y}Se$ single crystals, which have high $T_c \approx 40$ K and share similar Fermi surface topology as $A_xFe_{2-y}Se_2$, but are free from the sample complications. We found that the ambient-pressure $T_c \approx 41$ K is suppressed gradually to below 2 K upon increasing pressure to $P_c \approx 5$ GPa, above which a SC-II phase with higher $T_c$ emerges and the $T_c$ increases progressively to above 50 K up to 12.5 GPa. Interestingly, our high-precision resistivity data enable us to uncover the sharp transition of the normal state from a Fermi liquid for SC-I phase ($0 < P < 5$ GPa) to a non-Fermi-liquid for SC-II phase ($P > 5$GPa). In addition, the reemergence of high-$T_c$ SC-II phase is found to accompany with a concurrent enhancement of electron carrier density. Since high-pressure structural study based on the synchrotron X-ray diffraction rules out the structural transition below 10 GPa, the observed SC-II phase with enhanced carrier density should be ascribed to an electronic origin associated with a pressure-induced Fermi surface reconstruction.

# Introduction

Among the iron-based superconductors, the structural simplest FeSe and its derived materials have attracted tremendous attention recently due to its peculiar electronic properties and the great tunability of the superconducting transition temperature $T_c$. The bulk FeSe displays a relatively low $T_c \approx 8.5$ K within the peculiar nonmagnetic nematic phase below $T_s \approx 90$ K.[1] By intercalating some alkali-metal ions, ammonia, or organic molecules in between the adjacent FeSe layers, such as in $A_xFe_{2-y}Se_2$ (A = K, Rb, Cs, Tl),[2,3] $A_x(NH_3)_yFeSe$,[4] and (Li,Fe)OHFeSe,[5,6] high-$T_c$ superconductivity with $T_c$ above 30-40 K have been successfully achieved. More surprisingly, when a single unit-cell FeSe film is fabricated on the $SrTiO_3$ substrate, its $T_c$ can be raised up to 65-100 K.[7,8] Here, we refer these high-$T_c$ superconductors derived directly from FeSe as the SC-I phase. The superconducting mechanism for these SC-I phases has been subjected to extensive investigations, and the observed common Fermi surface topology consisting of only electron pocket in the Brillouin zone corners suggests that the electron doping plays an essential role for achieving high $T_c$,[9-11] in agreement with the gate-voltage regulation experiments on the FeSe flakes.[12]

Starting from the SC-I phase in $A_xFe_{2-y}Se_2$, Sun *et al.*[13] had reported in 2012 a pressure-induced sudden reemergence of a second superconducting phase (denoted as SC-II hereafter) with higher $T_c$ up to 48.7 K above ~ 10 GPa. A similar pressure-induced SC-II phase has also been observed in $Cs_{0.4}(NH_3)_yFeSe$ under high pressure.[14] Although the reemergence of SC-II phase with higher $T_c$ is quite intriguing and different pairing symmetry has been proposed theoretically,[15] the intrinsic superconducting- and normal-state properties have been poorly characterized so far due to some sample and technical difficulties. For example, $A_xFe_{2-y}Se_2$ superconductors are prone to phase separation accompanied with the intergrowth of antiferromagnetic insulating $A_2Fe_4Se_5$ phase.[16] In addition, only polycrystalline samples have been studied under pressure for $Cs_{0.4}(NH_3)_yFeSe$ which is extremely sensitive to air.[14] Moreover, high-pressure technique capable of both large pressure capacity ***and*** good hydrostaticity is required in order to obtain reliable superconducting and normal-state properties. Therefore, these complexities have hampered a proper understanding on the intriguing SC-II phase of these FeSe-derived systems.

In order to approach this intriguing problem, we turn our attention to the recently discovered $(Li_{1-x}Fe_x)OHFe_{1-y}Se$,[5,6] which is free from phase separation, relatively stable in air, and more importantly, can be obtained in high quality single-crystal form via a specially designed hydrothermal ion-exchange method.[17] (Li$_{0.84}$Fe$_{0.16}$)OHFeSe with an optimal $T_c \approx 41$ K is heavily electron doped having only electron pockets at the corner of Brillouin zone, similar as A$_x$Fe$_{2-y}$Se$_2$ and monolayer FeSe/SrTiO$_3$ film.[10,18] Since a high-pressure study on (Li$_{1-x}$Fe$_x$)OHFe$_{1-y}$Se has not been reported so far, the following questions remain open: (i) how will the SC-I phase response to pressure? and (ii) will a SC-II phase also emerge under pressure as a common phenomenon? Moreover, the distance between two adjacent FeSe layers in (Li$_{1-x}$Fe$_x$)OHFe$_{1-y}$Se is much larger than that in bulk FeSe and A$_x$Fe$_{2-y}$Se$_2$, which signals a weak interlayer interaction and an enhanced two-dimensional nature of the electronic structure.[19] It thus has been considered as a better proxy of the monolayer FeSe film but is more stable and free from interface effects.[10] These factors together make it indispensable to perform a high-pressure study on (Li$_{1-x}$Fe$_x$)OHFe$_{1-y}$Se.

Here we performed detailed magnetotransport measurements on the (Li$_{1-x}$Fe$_x$)OHFe$_{1-y}$Se single crystals under hydrostatic pressures up to 12.5 GPa with a cubic anvil cell apparatus.[20] The main results are summarized in Fig. 1. We found that the ambient-pressure SC-I phase is suppressed gradually with increasing pressure to $P_c \approx 5$ GPa, above which a new SC-II phase with higher $T_c$ over 50 K emerges gradually. Importantly, our high-precision resistivity data enable us to uncover a sharp transition of the normal state from a Fermi liquid for SC-I phase to a non-Fermi-liquid for SC-II phase. In addition, the reemergence of higher $T_c$ SC-II phase is found to accompany with a concurrent enhancement of electron carrier density. Such information was unavailable in all previous high pressure studies on the FeSe-derived superconductors. The present work thus provides positive correlations between the high-$T_c$ superconductivity in SC-II with a Fermi surface reconstruction, which is not induced by a structural transition as confirmed by our high-pressure structural study.

## Experiment details

(Li$_{1-x}$Fe$_x$)OHFe$_{1-y}$Se single crystals used in the present study were grown with a hydrothermal ion-exchange technique by using a large insulating K$_{0.8}$Fe$_{1.6}$Se$_2$ crystal as a matrix. Details about the crystal growth and sample characterizations at ambient pressure can be found in the previous studies.[17] High-pressure transport and ac magnetic susceptibility were performed in the palm cubic anvil cell (CAC) apparatus.[20] The standard four-probe method was employed for resistivity measurement with the current applied within the *ab* plane and the magnetic field along the *c* axis. The $\rho_{xy}(H)$ and $\rho_{xx}(H)$ data were anti-symmetrized (symmetrized) with respect to the magnetic field between +5 and -5 T. Glycerol was employed as the pressure transmitting medium. The pressure values inside the CAC were calibrated at room temperature by observing the characteristic transitions of bismuth. The mutual induction method was used for the ac magnetic susceptibility measurements. High-pressure synchrotron X-ray diffraction (SXRD) was also measured at room temperature with diamond anvil cell (DAC) in order to investigate the structural changes under pressure. Glycerol was also used as the pressure medium. The pressure in DAC was monitored with the typical ruby fluorescence method.

## Results and discussions

Fig. 2(a) shows the temperature dependence of resistivity $\rho(T)$ for a $(Li_{1-x}Fe_x)OHFe_{1-y}Se$ single crystal ($T_c \approx 41$ K at ambient pressure) measured under various hydrostatic pressures up to 12.5 GPa in the whole temperature range. As can be seen, $\rho(T)$ in the normal state first decreases significantly and then becomes nearly unchanged above 6.5 GPa; the broad hump feature at high temperature also smears out gradually upon increasing pressure. The superconducting $T_c$ displays a non-monotonic variation with pressure, which can be seen more clearly from the vertically shifted $\rho(T)$ data below 100 K as shown in Fig. 2(b). Here we define the onset $T_c^{onset}$ (down-pointing arrow) as the temperature where $\rho(T)$ starts to deviate from the extrapolated normal-state behavior, and determine $T_c^{zero}$ (up-pointing arrow) as the zero-resistivity temperature. As can be seen, upon increasing pressure to 5 GPa, $T_c^{onset}$ is suppressed gradually to ~13 K and $T_c^{zero}$ can hardly be defined down to 1.4 K, the lowest temperature in our present study. Interestingly, when increasing pressure to 6.5 GPa, a broad superconducting transition appears again with the $T_c^{onset}$ raised to ~31 K and $T_c^{zero}$ at ~12 K, thus evidencing the emergence of the SC-II phase. With further increasing pressure, both $T_c^{onset}$ and $T_c^{zero}$ move up progressively and the superconducting transition becomes sharper. Finally, $T_c^{onset}$ and $T_c^{zero}$ reach 52.7 and 46.2 K, respectively, at $P_{max}$ = 12.5 GPa. A closer inspection of the $\rho(T)$ data in Fig. 2(b) also reveals a gradual evolution of the temperature dependence of normal-state resistivity under pressure, which will be discussed in detailed below.

The superconducting transitions have been further verified by the ac magnetic susceptibility $4\pi\chi(T)$ shown in Fig. 2(c), in which the superconducting diamagnetic signal appears below $T_c^\chi$ as indicated by the arrows. The obtained $T_c^\chi$ first decreases with pressure, reverses the trend near $P_c \approx 5$ GPa, and then increases quickly with further increasing pressure, in well agreement with resistivity data. In addition, the transition in $4\pi\chi(T)$ is broad when the resistivity transition is broad for $5 < P < 8$ GPa. Nevertheless, the superconducting shielding volume reaching over 60-70% confirmed the bulk nature of the observed superconductivity in both SC-I and SC-II phases.

The pressure dependences of the obtained $T_c^{onset}$, $T_c^{zero}$, and $T_c^\chi$ for the studied $(Li_{1-x}Fe_x)OHFe_{1-y}Se$ are displayed in Fig. 1(a), which evidenced explicitly the gradual suppression of the SC-I phase followed by the reemergence of the SC-II phase above the critical pressure $P_c \approx 5$ GPa. It looks that the SC-II phase will exhibit a dome-shaped $T_c(P)$ with the maximum taking place around 12-13 GPa. It is interesting to note that in the SC-I region $T_c^\chi$ agrees well with $T_c^{zero}$ as commonly seen in most superconductors, whereas in the SC-II region $T_c^\chi$ follows the $T_c^{onset}$, implying that a considerable superconducting volume already appears near $T_c^{onset}$ despite of a broad transition. Although the observation of pressure-induced SC-II phase in $(Li_{1-x}Fe_x)OHFe_{1-y}Se$ in the present study is qualitatively similar with those reported in $A_xFe_{2-y}Se_2$ and $Cs_{0.4}(NH_3)_yFeSe$,[13,14] there are some quantitative differences in comparison with those previously studies: (i) the obtained $T_c^{onset}$ here is higher, exceeding 50 K for the first time; (ii) $T_c^{zero}$ that has never been achieved for the SC-II phase in the previous studies using DAC is successfully reached here due to a better sample quality and improved hydrostaticity in CAC; (iii) the SC-II phase appears gradually and exists in a wide pressure range. We have measured another $(Li_{-1-x}Fe_x)OHFe_{1-y}Se$ sample with a lower carrier density and thus a lower $T_c \approx 28$ K at ambient-pressure and observed very similar behaviors featured by two superconducting domes separated at a lower critical pressure of $P_c \approx 3$ GPa. Details can be found in the Fig. S1 of Supplementary Materials. These experiments thus confirm that the pressure-induced reemergence of SC-II phase is likely a universal phenomenon in the $(Li_{-1-x}Fe_x)OHFe_{1-y}Se$ system, or even in the FeSe-derived high-$T_c$ superconductors taking together the previous studies.

To uncover the origin of such an intriguing phenomena, experimentally we need to first characterize the normal-state properties which are usually tightly correlated with the superconducting states for the unconventional superconductors. A distinct change on the temperature dependence of normal-state $\rho(T)$ has already been noticed in Fig. 2(b). To quantify this evolution, we display the $\rho(T)$ data in a double-logarithmic plot of $\log(\rho-\rho_0)$ versus $\log T$ in Fig. 3(a), where $\rho_0$ is the residual resistivity at zero temperature. The slope of these curves corresponds to the resistivity exponent $n$ in $\rho \propto T^n$, which evolves from a Fermi-liquid $n = 2$ for $0.7 \leq P \leq 4$ GPa, through some intermediate $2 < n < 1.5$ for $P = 5$ and 6.5 GPa, and finally to non-Fermi-liquid $n \leq 1.5$ for $P > 6.5$ GPa. Such an evolution can be visualized more profoundly in a contour plot of the resistivity exponent $n \equiv \mathrm{dlog}(\rho-\rho_0)/\mathrm{dlog}T$ superimposed in Fig. 1(a). The observed sharp transition of the normal-state behavior thus signals distinct superconducting states for the SC-I and SC-II phases. In particular, the nearly linear-in-$T$ behavior for the SC-II phase resembles those of the optimal doped cuprates and iron-pnictides superconductors, thus implying an unconventional mechanism for the emergent SC-II phase.[21] We want to underline that our high-precision resistivity data enable us to unveil the *non-Fermi-liquid* normal state of the SC-II phase for the first time.

In order to gain further insights into the peculiar non-Fermi-liquid behavior of SC-II phase, we tried to probe the electronic structure information via measurements of magnetoresistance (MR) and Hall effect under pressure. Fig. 3(b) displays the field dependence of in-plane MR$(H) \equiv [\rho(H)/\rho(0)-1] \times 100\%$ and Hall resistivity $\rho_{xy}(H)$ in the normal state just above $T_c$ under various pressures up to 8 GPa. As can be seen, the MR is small and decreases gradually from 3% at 0.7 GPa to below 0.5% at 8 GPa. All $\rho_{xy}(H)$ curves exhibit a linear-in-$H$ behavior with a negative slope, signaling that the electron-type carriers dominate the charge transport in both the SC-I and SC-II phases. In contrast with the monotonic decrease of MR, $\rho_{xy}$ displays a non-monotonic variation with pressure. Here, we obtained the Hall coefficient $R_H \equiv \mathrm{d}\rho_{xy}/\mathrm{d}H$ as the slope of a linear fitting to $\rho_{xy}(H)$, and plotted the field dependence of $R_H(H)$ in Fig. 1(b). As can be seen, $R_H$ is negative, and its magnitude first increases slightly with pressure and then experiences a quick reduction above 4 GPa. Assuming a simple one-band contribution, the electron-type carrier density can be estimated as $n_e = -1/(R_H \cdot e)$. As shown in Fig. 1(b), $n_e$ takes a relatively constant value of $\sim 2 \times 10^{27}$ m$^{-3}$ within the SC-I region for $P < 5$ GPa, above which it increases linearly to a large value of $\sim 9 \times 10^{27}$ m$^{-3}$ at 8 GPa, tracking nicely the trend of $T_c(P)$. These results demonstrate that the emergence of SC-II phase with higher $T_c$ is accompanied with a concurrent enhancement of electron carrier density. Such a positive correlation between $T_c$ and $n_e$ is not surprising in the FeSe-based superconductors as mentioned above, but the origin of the pressure-induced enhancement of $n_e$ in the SC-II phase deserves in-depth investigations.

To this end, we first checked if a structural transition takes place near $P_c \approx 5$ GPa. Fig. 4(a) displays the high-pressure SXRD patterns of $(Li_{1-x}Fe_x)OHFe_{1-y}Se$ measured at room temperature up to 14 GPa. Except for a small peak near ~12° from the gasket (marked by asterisk), all the peaks can be indexed in the tetragonal *P4/nmm* (No. 129) space group. As can be seen, no obvious structural transition can be discerned in the investigated pressure range. The relative peak intensities are altered when applying pressure above 0.8 GPa due to the development of preferred orientation, as exemplified by the (200) peak near 20°. We have applied the Rietveld refinements on these SXRD patterns and extracted the unit-cell parameters as a function of pressure as depicted in Fig. 4(b-d). As can be seen, both the lattice parameters $a$, $c$, and the unit-cell volume $V$ decrease smoothly up to 10 GPa, above which $a$ tends to level off and $c$ experiences a steep drop. Since the SXRD peaks become relatively broad above 10 GPa due to the solidification of liquid pressure medium, the anomalous structure changes above 10 GPa deserves further studies with better resolved SXRD patterns by employing the gas pressure medium. But, our present high-pressure structural study rules out any structural transition below 10 GPa as the possible cause for the observed enhancement of carrier density and the emergence of SC-II phase.

Without a structural transition taking place at $P_c \approx 5$ GPa, we are left with a pure electronic origin for the observed SC-II phase. Unfortunately, a direct probe of the electronic structure near Fermi surface (FS), *e.g.* with APRES, is impossible under high pressure. Recently, a second high-$T_c$ dome and a second enhancement of superconductivity were reported in the heavily K-deposed FeSe film grown on SiC [22] and the $SrTiO_3$ substrates[23], respectively. By taking advantage of the in-situ APRES measurements, the second enhancement of superconductivity in the latter case has been attributed to a Lifshitz transition associated with the emergence of an electron pocket at the Γ point of Brillouin zone center.[23] Since no extra electron carriers were purposely doped into $(Li_{1-x}Fe_x)OHFe_{1-y}Se$ in the present case, the dramatic enhancement of carrier density $n_e$ and $T_c$ above 5 GPa cannot be ascribed to a doping-induced Lifshitz transition. In addition, a further charge transfer from the insulating (Li,Fe)OH layer to the FeSe layer is unlikely since the reemergence of SC-II phase has been observed universally in different classes of FeSe-derived systems. Instead, a Lifshitz transition in the FeSe layers might take place via a pressure-induced Fermi surface reconstruction. Recent scanning tunneling spectroscopy study on the $(Li_{1-x}Fe_x)OHFeSe$ single crystal has identified two electron pockets at the M point associated with the $d_{xy}$ and $d_{xz}/d_{yz}$ orbitals, respectively.[24] The observed ($\pi$, 0.67$\pi$) wave vector in the spin resonance spectroscopy with inelastic neutron-scattering is consistent with the nesting vector between the 2D electron Fermi pockets[25]. Whether these electron pockets at M point undergo reconstruction or another electron pockets emerge near $\Gamma$ point deserve further theoretical studies. In any case, the reemergence of higher $T_c$ SC-II phase developed from the unusual non-Fermi-liquid normal states with enhanced electronic charge carriers outlines important constrains for further investigations.

Finally, it is noteworthy that the normal-state resistivity of the cuprate superconductors, *e.g.* the overdoped $La_{2-x}Sr_xCuO_4$ and $La_{2-x}Ce_xCuO_4$,[26,27] behaves as $\rho(T) \sim T^{1.6}$ at the verge of the superconducting dome, which has been attributed to quantum criticality. The observation of similar power-law behavior near the border of SC-II dome in the present $(Li_{1-x}Fe_x)OHFe_{1-y}Se$ thus points to the plausible common physics that awaits for in-depth explorations in future.

# Conclusion

In summary, we have measured the resistivity of $(Li_{1-x}Fe_x)OHFe_{1-y}Se$ single crystal under hydrostatic pressures up to 12.5 GPa with a cubic anvil cell apparatus, and observed a gradual suppression of superconductivity followed by reemergence of a second high-$T_c$ superconducting (SC-II) phase above $P_c \approx 5$ GPa. The highest $T_c$ reaches ~52 K, which is highest among the bulk form of FeSe-derived superconductors. The SC-II phase is confirmed to develop from a peculiar non-Fermi-liquid normal state featured by dominant electron-type charge carriers and enhanced carrier density. Since no any structural transition was detected below 10 GPa, the observation of SC-II phase with enhanced carrier density has been ascribed to an electronic origin associated with Fermi surface remonstration.

# Acknowledgements

This work is supported by the National Science Foundation of China (Grant No. 11574377, 11574370, U1530402), the National Basic Research Program of China (Grant Nos. 2014CB921500, 2017YFA0303000, and 2016YFA0300301), the Strategic Priority Research Program and Key Research Program of Frontier Sciences of the Chinese Academy of Sciences (Grant Nos. XDB07020100, QYZDB-SSW-SLH013, QYZDY-SSW-SLH001 and QYZDY-SSW-SLH008). The XRD measurements were performed at the BL15U1 beamline,

Shanghai Synchrotron Radiation Facility (SSRF) in China. The authors would like to thank Drs. Aiguo Li and Ke Yang for beamline technical support.

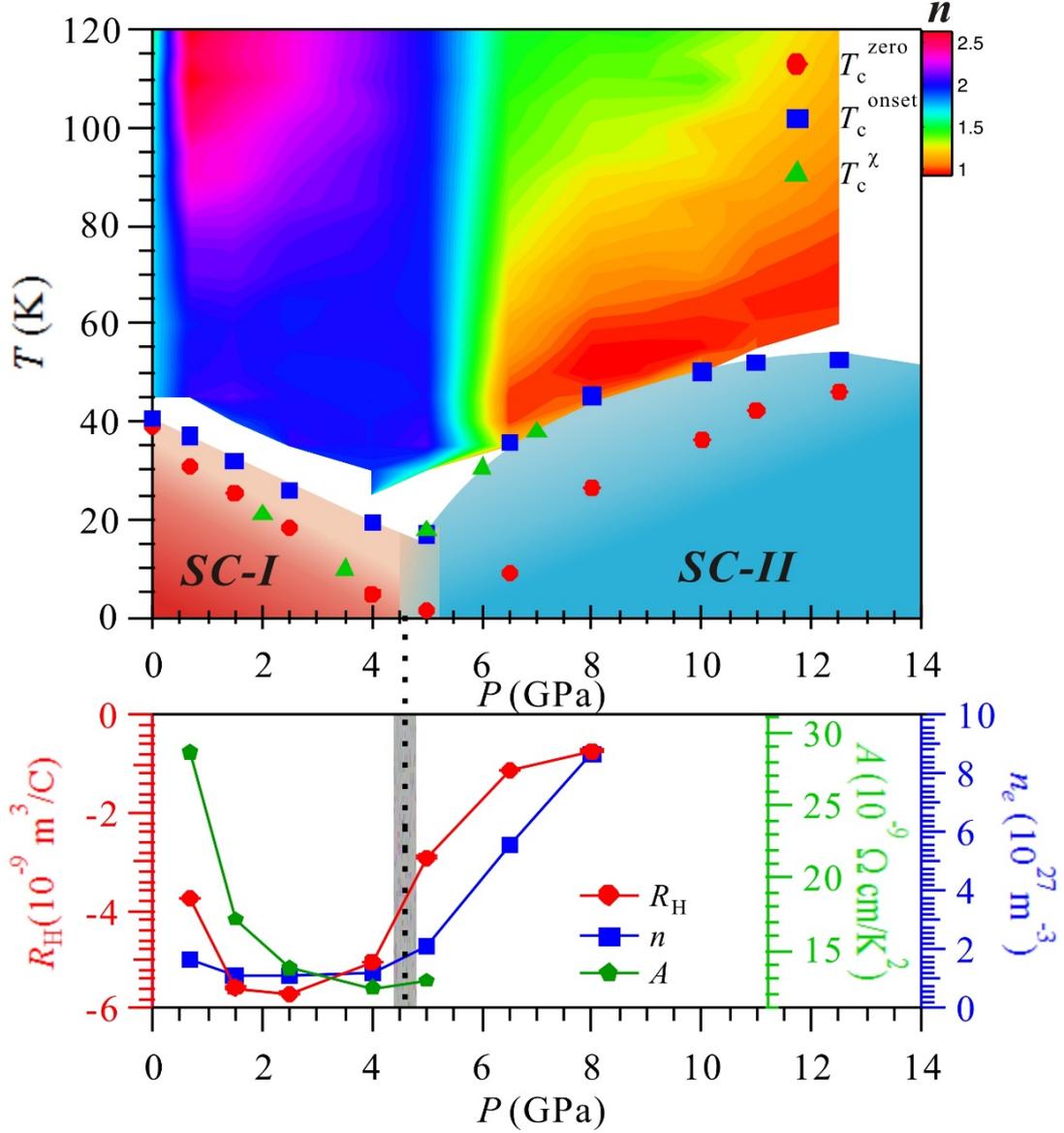

**Fig. 1** T-P phase diagram of $(Li_{1-x}Fe_x)OHFe_{1-y}Se$ single crystal (x ≈ 0.84, $T_c$ = 40 K at ambient pressure). (a) Pressure dependence of the superconducting transition temperatures $T_c$s and a contour color plot of the normal-state resistivity exponent $n$ up to 12.5 GPa. The values of $T_c^{onset}$, $T_c^{zero}$, and $T_c^{\chi}$ were determined from the high-pressure resistivity and ac susceptibility shown in Fig. 2. The temperature dependence of $n$ are extracted from $d\ln(\rho-\rho_0)/d\ln T$ for each pressure shown in Fig. 3(a). (b) Pressure dependences of the Hall coefficient $R_H$ and the electron density $n_e$ up to 8 GPa extracted from the transverse resistivity $\rho_{xy}$ shown in Fig. 3(c). Pressure dependence of the resistivity coefficient $A$ in the plot of $\rho \sim AT^2$ below 5 GPa. A double-domed $T_c(P)$ accompanied with distinct normal-state properties for each superconducting phase is clearly observed. The reemergence of the SC-II phase with higher $T_c$ is accompanied with a dramatic enhancement of the carrier density.

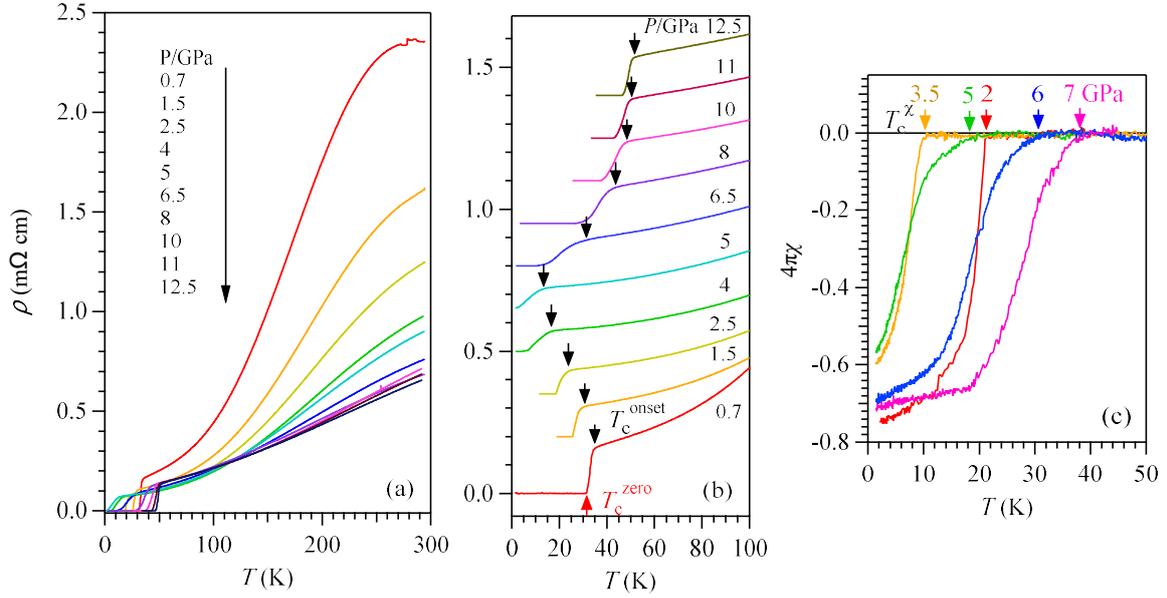

Fig. 2 High-pressure resistivity $\rho(T)$ and ac magnetic susceptibility $4\pi\chi(T)$ for $(Li_{1-x}Fe_x)OHFe_{1-y}Se$ single crystal. (a) $\rho(T)$ curves in the whole temperature range illustrating the overall behaviors under pressure up to 12.5 GPa. (b) $\rho(T)$ curves below 100 K illustrating the variation with pressure of the superconducting transition temperatures. Except for data at 0.7 GPa, all other curves in (b) have been vertically shifted for clarity. The onset $T_c^{onset}$ (down-pointing arrow) was determined as the temperature where resistivity starts to deviate from the extrapolated normal-state behavior, while the $T_c^{zero}$ (up-pointing arrow) was determined as the zero-resistivity temperature. (c) $4\pi\chi(T)$ curves measured under different pressures up to 7 GPa. The superconducting diamagnetic signal appears below $T_c^\chi$.

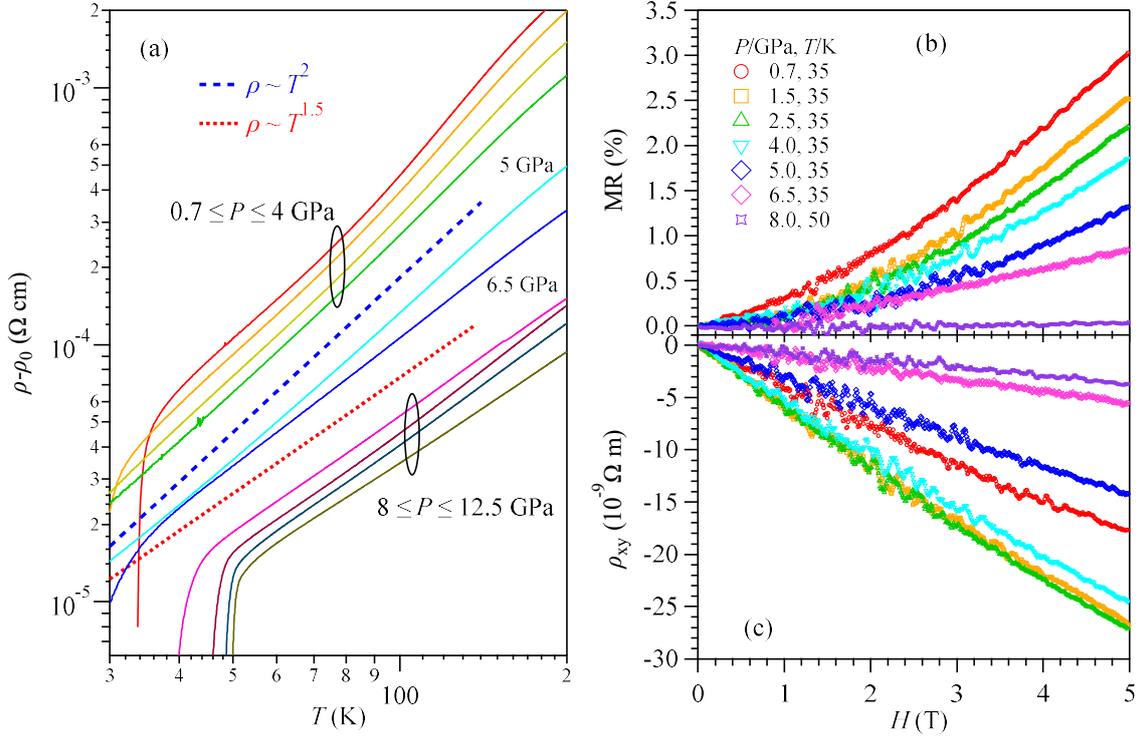

Fig. 3 Normal-state transport properties of $(Li_{1-x}Fe_x)OHFe_{1-y}Se$ under high pressure. (a) A double logarithmic plot of $(\rho-\rho_0)$ versus $T$ illustrating the variation with pressure of the normal-state resistivity from the Fermi-liquid $\rho \sim T^2$ for $P < 5$ GPa to non-Fermi-liquid $\rho \sim T^{1.5}$ behavior for $P > 6.5$ GPa. Except for the curve at 5 GPa, all other curves have been vertically shifted for clarity. Field dependence of (b) the magnetoresistivity MR and (c) the transverse resistivity $\rho_{xy}$ at the normal state just above $T_c$ under various pressures. The Hall coefficient $R_H$ are determined from the field derivative of $\rho_{xy}$, $R_H \equiv d\rho_{xy}/dH$, at each pressure as shown by the solid lines in (c).

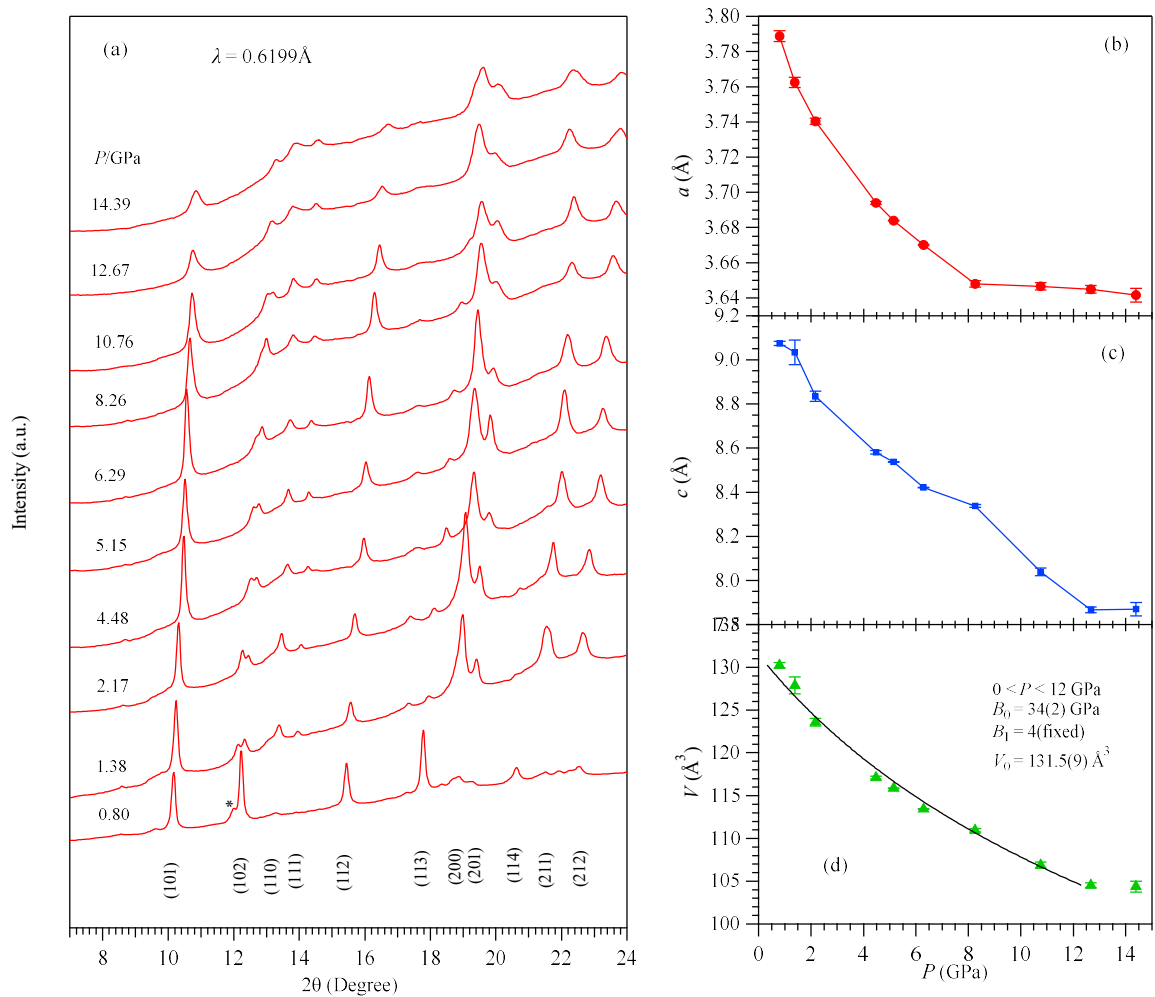

Fig. 4 High-pressure synchrotron XRD. (a) XRD patterns of $(Li_{1-x}Fe_x)OHFe_{1-y}Se$, and (b-d) pressure dependence of unit-cell parameters $a$, $c$, and $V$. The solid lines in (d) are the Birch-Murnaghan fitting curves used to extract the bulk modulus given in the inset.